# Modification of band alignment at interface of $Al_yGa_{1-y}Sb/Al_xGa_{1-x}As$ type-II quantum dots by concentrated sunlight in intermediate band solar cells with separated absorption and depletion regions

A. Kechiantz[a]*, A. Afanasev[a], J.-L. Lazzari[b]

[a]Dept of Physics, The George Washington Univ., 725 21st Street, NW, Washington, DC, USA, 20052, kechiantz@gwu.edu, afanas@gwu.edu ; [b]Centre Interdisciplinaire de Nanoscience de Marseille, CINaM, UMR 7325 CNRS – Aix-Marseille Université, Campus de Luminy, Case 913, 163 Avenue de Luminy, 13288 Marseille Cedex 9, France; lazzari@cinam.univ-mrs.fr; *On leave: Institute of Radiophysics and Electronics, National Academy of Sciences, 1 Brothers Alikhanyan st., Ashtarak 0203, Armenia

## ABSTRACT

We propose a new intermediate band GaAs solar cell comprising an $Al_xGa_{1-x}As$ absorber with built-in GaSb type-II quantum dots (QDs) [a gradual $Al_xGa_{1-x}As$ absorber with built-in $Al_yGa_{1-y}Sb$ QDs *(0<x=y<0.40)* as a variant] separated from the depletion region. We study the modification of the band alignment at type-II interface by two-photon absorption of concentrated sunlight. Our calculation shows that photogenerated carriers produce localized exciton-like electron-hole pairs spatially separated at QDs. Local field of such pairs may essentially modify potential barrier surrounding QDs, increase recombination lifetime of mobile carriers and additional photocurrent generated by two-photon absorption. Concentration of about 300-sun pushes by 15% up the conversion efficiency as compared to the efficiency of the reference single junction GaAs solar cell without QDs.

## 1 INTRODUCTION

The intermediate band (IB) is a concept introduced by Luque and Marti during 1997 in the realm of solar cells[1]. This concept gives a way for putting into use energy of sub-band gap photons, in fact, due to the non-linear effect of two-photon absorption[2]. If the total energy of two sub-band-gap photons exceeds the energy of band-gap, consecutive absorption of two photons may transfer a valence band electron into the conduction band resulting in an additional photocurrent. As known, both concentration of light and resonant absorption escalates non-linear optical effects. The IB concept exploits a band of intermediate electronic states located in the semiconductor band gap for resonant absorption. If IB-states would not result in electron-hole recombination (like impurity defect states), IB solar cells would convert up to 63% of concentrated sunlight into electricity[3].

The IB concept does not identify a design for IB solar cells but it gives interesting prospects to tailor IB-states and the cell[4]. While these developments are still in progress, a few experiments have already been performed with quantum dots (QDs) [5, 6]. QDs are a family of zero-dimensional semiconductor building-blocks that exhibits unique electronic properties. The confined electronic states of type-I QDs were used to produce IB-states located within the depletion layer of a single p-n-junction solar cell. The experiments with such IB solar cells have shown that type-I QDs facilitates two-photon absorption of sub-band gap photons, however, they also lead to generation of additional dark current reducing both open circuit voltage and conversion efficiency of the cells. Our theoretical[7] study confirmed that type-I lineup of energy bands enable these QDs easily to capture electrons from the conduction band and holes from the valence band, so that type-I QDs present themselves as recombination centers. Since the depletion region is the most sensitive part of solar cells where electronic states easily facilitate recombination, we inferred that location of QDs (especially type-I QDs) "inside" the depletion region boosted the dark current of QD IB solar cells.

Type-II QDs are another family of zero-dimensional semiconductor building-blocks that also exhibits unique electronic properties. For instance, type-II QDs, whose confined electronic states are in the valence band, spatially separate mobile electrons of the conduction band from holes confined in QDs. Such separation may decrease recombination rate of these carriers. In GaSb/GaAs strained semiconductor systems lifetime of mobile electrons with respect to recombination with holes confined in GaSb type-II QDs is about $10 ns$[8], which is very close to the electron-hole radiative recombination lifetime in GaAs. We have used this unique property in our calculation[2, 7, 9] to show that IB composed of confined electronic states of type-II QDs located "outside" of the depletion region may have no effect on the electron-hole overall recombination rate and hence the dark current. This gives more flexibility for designing the QD IB solar cell structures. The large offset, direct band gaps, type-II (staggered) misalignment of the conduction and valence bands, and well-developed fabrication technology make GaAs/GaSb strained system a good subject for studying of potential of unique electronic properties of QDs for IB applications. Here we focus on high density of confined electronic states in QDs. These states are discrete and they spread from the valence band edge deep into the semiconductor band gap. Potential barriers around QDs are highly sensitive to the occupation of such confined states.

In this paper we study a new IB solar cell concept comprising an $Al_xGa_{1-x}As$ absorber with built-in GaSb type-II QDs [a gradual $Al_xGa_{1-x}As$ absorber with built-in $Al_yGa_{1-y}Sb$ QDs *(0<x=y<0.40)* as a variant] located "outside" the depletion region in the p-doped part of a GaAs p-n-junction. The focus is on modification of the band alignment at QDs by two-photon absorption of concentrated sunlight and its correlation with the IB solar cell performance. Our calculation of additional photocurrent generated in GaSb/GaAs type-II QD IB solar cell due to two-photon absorption of sub-band gap photons shows that 300-sun concentration of light may push by 15% up the conversion efficiency as compared to the efficiency of the reference single junction GaAs solar cell without QDs.

## 2 DESIGN

### 2.1 Electronic features of the type-II QD IB solar cell design

The key element of type-II QD IB solar cell studied here is the type-II QD absorber of sub-band gap photons. Since electronic states located in the depletion region facilitate generation of recombination dark current, resulting in reduction of the open circuit voltage, we believe that the proper position of the QD absorber is outside the depletion region as shown in Figure 1. In our proposed design of the IB solar cell, the QD absorber is sandwiched between a thin $p^+$-doped $Al_xGa_{1-x}As$ buffer layer grown on an $n^+$-doped GaAs substrate and a $p^+$-doped $Al_xGa_{1-x}As$ cap layer. The QD absorber is an epitaxial stack comprising GaSb strained QD layers alternating with p-doped $Al_xGa_{1-x}As$ spacers. All layers of the stack are within the electron diffusion length distance from the depletion region. The spacers compose non-tunneling barriers surrounding QDs in the valence band. The buffer and the substrate compose an ideal p-n-junction. The buffer layer is thick, for instance 200nm, to separate the edge of depletion region from the absorber and prevent electron tunneling through the p-n-junction into electronic states confined in QDs. Another feature of the cell design shown in Figure 1 is doping of the stack and QDs by the $p^+$-doped $Al_xGa_{1-x}As$ cap and buffer layers. Such doping of the lower doped layers by carriers from the higher doped layers, known as modulation doping, equalizes Fermi level across the cell.

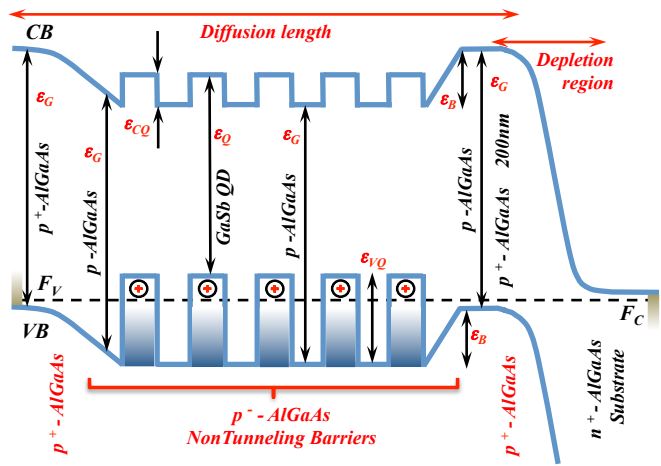

Figure 1. Energy band diagram of GaSb/AlGaAs type-II QD IB solar cell. The stack of GaSb QDs is embedded in $p^+$-doped $Al_xGa_{1-x}As$ layer so that the stack does not touch the depletion region. $Al_xGa_{1-x}As$ barriers in valence band are thick to prevent tunneling for holes confined in QDs.

The modulation doping lowers the conduction band edge of the stack relative to that in p⁺-doped $Al_xGa_{1-x}As$ cap and buffer layers by $\epsilon_B$ as shown in Figure 1. This reduction does not reach the depletion region of p-n-junction so that photoelectrons generated in the stack face blocking barrier $\epsilon_B$ in conduction band on their way to the p-n-junction. Due to the type-II misalignment of energy bands, conduction band electrons face also $\epsilon_{CQ}$ offset-barrier that spatially separates them from holes confined in QDs. The spatial separation slows down the lifetime associated with non-radiative inter-band recombination of mobile electrons with confined holes, for instance, to $10\ ns$ in GaSb/GaAs strained type-II QD system[8]. While the $\epsilon_{CQ}$ offset-barriers protect electrons from recombination with confined holes, they do not limit electron diffusion in conduction band of $Al_xGa_{1-x}As$ spacers as shown in Figure 2.

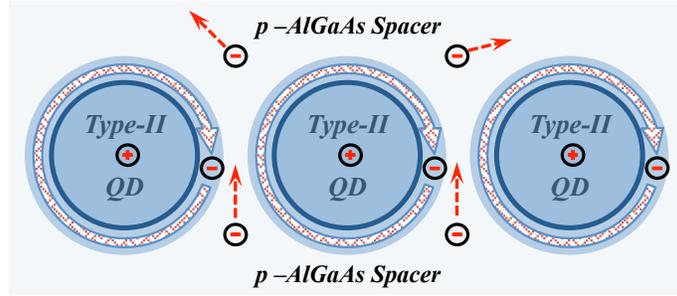

Figure 2. A schematic picture of the conduction band in the stack of GaSb/GaAs QD layers for QD absorber of type-II QD IB solar cell. GaSb QDs are sandwiched between p-doped $Al_xGa_{1-x}As$ spacer layers so that electrons can pass between QDs.

The volume of QDs is about $10^{-18}cm^3$. Such a small volume impacts the intra-band relaxation time in QDs. For instance, absorption of single sub-band gap photon injects a photoelectron from the valence band into the QD confined electronic state. Such absorption pushes up to $10^{18}cm^{-3}$ the local density of photoelectrons confined in the valence band by strained GaSb/GaAs type-II QD. Another sub-band gap photon may transfer this confined photoelectron into the conduction band, or the photoelectron may relax back into valence band by recombining (annihilation) with a mobile hole. Whichever is preferred depends on intensity (concentration) of sub-band gap photons and availability of mobile holes for annihilation. The $Al_xGa_{1-x}As$ separator layers create $\epsilon_{VQ}$ high offset-barrier around GaSb QDs as shown in Figure 1. These barriers spatially separate mobile holes from the confined states and reduce by $exp(\epsilon_{VQ}/kT)$ time the probability that a mobile hole can be found at the top of the valence band offset-barrier and hence its probability to enter into QDs. One can see that the density of mobile holes available for annihilation is essentially lower than $10^{18}cm^{-3}$ density of the confined photoelectrons. Then the density of mobile holes at the top of the offset-barrier limits the non-radiative intra-band annihilation in QDs by increasing the lifetime of photogenerated electrons confined in QDs $exp(\epsilon_{VQ}/kT)$ times.

It is well known that an absorption coefficient associated, for instance, with electron transition from the valence band into the conduction band is proportional to the density of electronic states occupied with electrons in the valence band and unoccupied states in the conduction band. In case of narrow energy band or single energy level within semiconductor band gap, absorption associated with electron transition from the valence band into the narrow band is proportional to the density of unoccupied confined electronic states in the narrow band while absorption associated with electron transition from the narrow band into the conduction band is proportional to the density of occupied confined electronic states in the narrow band. Since these two conditions seem incompatible, half occupation that results in matching of the quasi-Fermi level to the narrow band is the optimum for its assistance to the two-photon absorption of sub-band gap photons. In case of QDs, this condition is softened. A QD may have a set of discrete states within a wide energy range. For instance, a strained GaSb type-II QD comprises 15 confined electronic states in the valence band[10]. Their energy is spread over $300\ meV$ energy range in GaAs band gap as shown in Figure 1. Whichever state the quasi-Fermi level cross, the GaSb QD is in condition to facilitate the two-photon absorption of sub-band gap photons since a half of the state is occupied.

The small volume, $10^{-18} cm^3$, gives QDs one more unique property. It makes the local density of electronic states per QD extremely high since even a single confined state yields $10^{18} cm^{-3}$ density of states per QD. The strained GaSb type-II QD comprises 15 confined electronic states[10], which yields extremely high, $1.5 \times 10^{19} cm^{-3}$, local density of states. Since this density of states is very close to that in the valence band of GaSb bulk semiconductor, we can expect that the absorption coefficients per QD must be equal to that of direct band gap GaSb semiconductor, which is very favorable for the two-photon absorption of sub-band gap photons by the strained GaSb/GaAs type-II QD absorber.

## 2.2 Modification of the band edge alignment by concentrated sunlight

Concentrated sunlight rearranges distribution of charge carriers in the IB solar cell and split the Fermi level into quasi-Fermi levels for mobile electrons in conduction band, mobile holes in valence band, and photoelectrons confined in QDs. The modified energy band diagram is shown in Figure 3. The red dashed arrows denote the electron transfers to higher energy states due to photon absorption. The sub-band gap photon that transfers an electron in the valence band from the top of the $\epsilon_{VQ}$ offset-barrier into the states confined in QDs leaves a mobile hole at the top of the offset-barrier as shown in Figure 3. Such holes swiftly diffuse from QDs into the p$^+$-doped Al$_x$Ga$_{1-x}$As cap layer. A positive charge accumulated in the p$^+$-doped Al$_x$Ga$_{1-x}$As buffer layer balances the diffusion of generated holes into the buffer. The sub-band gap photon that transfers a confined electron from the valence band into the conduction band generates a mobile photoelectron on the top of the $\epsilon_{CQ}$ offset-barrier in QD as shown in Figure 3. Such photoelectrons swiftly escape from QDs, relax into the conduction band edge of Al$_x$Ga$_{1-x}$As spacers, and diffuse towards p-n-junction. Accumulation of these photoelectrons in the conduction band of the stack negatively charges the stack while accumulation of holes in the p$^+$-doped Al$_x$Ga$_{1-x}$As buffer layer positively charges the buffer. These charging lower the blocking barrier $\epsilon_B$ and enables generating photoelectrons pass through the depletion region as shown in Figure 3.

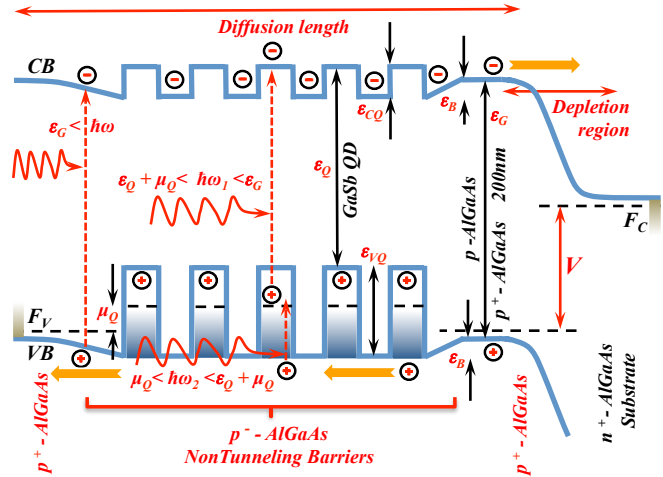

Figure 3. Energy band diagram of GaSb/GaAs type-II QD IB solar cell under concentrated sunlight illumination. The negative charge of photoelectrons accumulated in the stack and the positive charge accumulated in the buffer move up the conduction band edge in the stack, which reduces the blocking barrier $\epsilon_B$ as compared to that shown in Figure 1.

## 3 BALANCE OF CURRENTS

Assuming absorption of all incoming solar photons in each of $[\varepsilon_i, \varepsilon_j]$ spectral ranges, where $i$ and $j$ refer to either the conduction band, $C$, or the valence band, $V$, or QDs, $Q$, or the non-radiative electron transition, $N$, we used the principle of detailed balance to reduce currents generated by radiative electron transitions to $j_{ji} exp(\mu/kT)$, where $j_{ji}$ can be written as integral[2]

$$j_{ji} = \frac{2eX}{h^3 c^2 GF} \int_{\varepsilon_i}^{\varepsilon_j} \frac{exp[(-\mu)/kT]\varepsilon^2 d\varepsilon}{exp[(\varepsilon-\mu)/kT]-1} \tag{1}$$

In case of radiative recombination currents related to photon emission from the solar cell, $X = GF = 1$ and $\mu$ is the splitting of quasi-Fermi levels. In case of photocurrents related to solar photon absorption, $X$ is the concentration of solar light, $\mu = 0$, and $GF = 4.6 \times 10^4$ is the geometrical factor related to the angle that Earth is seen from Sun.

As it has been mentioned above, since the local density of both occupied and non-occupied states confined in QDs is very high, $\approx 10^{18} cm^{-3}$, the density of mobile photoelectrons in conduction band and mobile holes at the top of the $\epsilon_{VQ}$ offset-barrier around QDs determine currents generated in the stack by the allowed non-radiative electron transitions. Assuming the same quasi-Fermi level for conduction band electrons in both n- and p-doped sides of the p-n-junction, the non-radiative currents can be written as

$$j_{NQV}[1 - exp(-\mu/kT)] \tag{2}$$

$$j_{NCQ}[1 - exp\{(eV - \mu)/kT\}] \tag{3}$$

where $j_{NQV} = (e p_0 n_Q \Omega L/\tau_{ph}) exp(-\varepsilon_B/kT)$ and $j_{NCQ} = (e n_i^2 L/p_0 \tau_C) exp(\varepsilon_B/kT)$ refer to currents from the valence and conduction bands into QDs; $p_0$ is the concentration of holes in the cap and buffer $p^+$-doped layers; $n_i$ is the intrinsic concentration of carriers in Al$_x$Ga$_{1-x}$As spacer layers; $n_Q$ is the density of QDs; $\Omega$ is the volume of QD; $L$ is the thickness of the stack; $\tau_{ph}$ is the lifetime of intra-band relaxation (annihilation) in QDs due to optical phonons; $\tau_C$ is the lifetime of non-radiative inter-band recombination through QDs; $\mu$ is the separation of quasi-Fermi level in QD from that of the valence band; $kT$ is the temperature of the cell; $V$ is the bias voltage; $eV - \mu$ is the separation of the conduction band quasi-Fermi level from that of QDs; and $\varepsilon_B$ is the blocking barrier that limits the flow of holes into the stack from the cap and buffer layers. The charge accumulated in QDs, the p$^+$-doped Al$_x$Ga$_{1-x}$As buffer layer, and the conduction band in the stack determine the height $\varepsilon_B$ of this blocking barrier,

$$\varepsilon_B + \frac{4\sqrt{\varepsilon_B}(kT)^{3/2}}{e^2 l_D n_Q \Omega N_I L} + \frac{n_i^2 kT}{p_0 n_Q \Omega N_I} exp \frac{\varepsilon_B + eV}{kT} = \varepsilon_{VQ} - \mu - F_V \tag{4}$$

The photocurrent $j$ generated in QD IB solar cell consists of two components, which are the photocurrent $j_C$ generated due to absorption of the above-band gap photons directly in the conduction band and the additional photocurrent $j_Q$ generated due to the two photon absorption of sub-band gap photons in QDs, $j = j_C + j_Q$. Balance of $j_{VC}$ and $j_{CV}$ currents, $j_C = j_{VC} - j_{CV}$, yields the photocurrent $j_C$ the same expression as for the conventional photocurrent of the reference solar cell without QDs,

$$j_C = j_{SVC} - j_{CV}[exp(eV/kT) - 1] \tag{5}$$

Balance of $j_{SVQ}, j_{QV}$ and $j_{NQV}$ currents yields the net photocurrent, $j_Q = j_{SVQ} - j_{QV} - j_{NQV}$, into QDs from the valence band while balance of $j_{SQC}, j_{CQ}$ and $j_{NCQ}$ currents yields the net photocurrent, $j_Q = j_{SQC} - j_{CQ} - j_{NCQ}$, from QDs into the conduction band. Both net currents are equal to the additional photocurrent $j_Q$ generated in type-II QD IB solar cell,

$$j_Q = j_{SQC} - (j_{QC} + j_{NCQ})\left(exp\frac{eV-\mu}{kT} - 1\right) = j_{SVQ} - j_{VQ}\left(exp\frac{\mu}{kT} - 1\right) - j_{NQV}\left(1 - exp\frac{-\mu}{kT}\right) \tag{6}$$

Here $j_{NCQ} \sim exp(\varepsilon_B/kT)$ and $j_{NQV} \sim exp(-\varepsilon_B/kT)$. Substituting $j_C$ and $j_Q$ reduces the photocurrent $j = j_C + j_Q$ to

$$j = j_{SVC} + j_{SQC} - j_{CV}[exp(eV/kT) - 1] - (j_{QC} + j_{NCQ})\left(exp\frac{eV-\mu}{kT} - 1\right) \tag{7}$$

## 4 RESULTS AND DISCUSSION

The photocurrent generated in ideal IB solar cells has been calculated[3] assuming that only radiative electron transitions are allowed. Here we slightly soften this condition assuming that non-radiative electron transitions are allowed for inter-band recombination of the conduction band electrons with holes confined in QDs and for the intra-band annihilation of confined photoelectron with mobile hole in QDs, which is a capture of a mobile hole into a confined state of QD. Both non-radiative transitions increase the dark current. However, we have shown in previous papers[2, 7, 9] that when QDs are located outside the depletion region of type-II QD IB solar cell, the rate of these non-radiative transitions can be reduced.

Sub-bandgap photons generate mobile photoelectrons that swiftly escape from conduction band of few $nm$-thick QDs into the conduction band of $Al_xGa_{1-x}As$ spacers and relax there in $1 ps$. The corresponding holes remain strongly confined in QDs. The escaped photoelectrons diffuse towards the p-n-junction. If the stack comprises graded $Al_xGa_{1-x}As$ spacers, a drift driven by the pulling field of the graded spacers may enforce this diffusion so that photoelectrons become able to pass through the $500 nm$ stack in $50 ps$, which is much shorter than their inter-band recombination lifetime, $1 ns - 10 ns$[8].

Solution of Equations (1) - (7) yields photocurrent and hence all photovoltaic characteristics of p-doped GaSb/AlGaAs strained type-II QD IB solar cell. In particular, Figure 4 displays the conversion efficiency $\eta$ of this solar cell as a function of sunlight concentration X. Our calculation shows that concentration of sunlight produces higher cell performance for two reasons. First, it supports generation of additional photocurrent by the two-photon absorption of sub-band gap photons. Second, it lowers the blocking barrier, $\epsilon_B$, which blocks photoelectrons to reach the p-n-junction. The blocking barrier is highly sensitive to the charge accumulated in GaSb QDs and AlGaAs buffer. Since concentration of sunlight modifies the charge, it also modifies the blocking barrier. However, concentration of about 300-sun reduces the blocking barrier $\varepsilon_B$ of ideal GaSb/GaAs QD IB solar cell to the thermal energy of mobile carriers. Such a small barrier cannot limit photoelectron diffusion towards p-n-junction, therefore photovoltaic performance meet the Luque-Marti limit at 300-sun concentration as shown in Figure 4. Our calculation shows that concentration of sunlight from 1-sun to 500-sun raise the efficiency of proposed GaSb/GaAs QD IB solar cell from 30% to 50%. Further concentration of light has little effect.

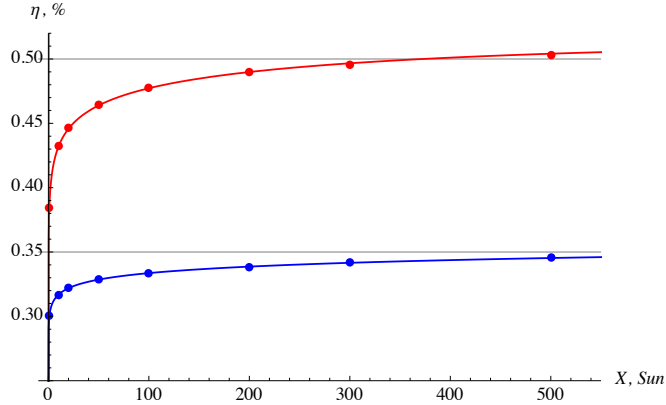

Figure 4. Conversion efficiency $\eta$ of p-doped GaSb/GaAs type-II QD IB solar cell as a function of sunlight concentration: radiative limit (red dots). Shockly-Queisser limit of reference GaAs solar cell (blue dots). Solid lines are $a + b \times ln(X)$ approximation of efficiency.

## 5 CONCLUSIONS

In conclusion our study shows that QDs may help generation of the additional photocurrent by resonant two-photon absorption of sunlight in p-doped type-II GaSb/GaAs QD IB solar cell. In the proposed cell design QDs are located outside the depletion region. Special caution should be taken here because, like artificial atoms, QDs may easily convert their confined ground electronic state into fast recombination level. Such conversion of QD states degrades both additional and conventional photocurrent of IB solar cells because confined states promptly achieve equilibrium with conduction or valence bands.

Our calculation shows that concentration of sunlight enforces two-photon absorption in proposed GaSb/GaAs QD IB solar cell. It also reduces recombination through the "outside" QDs, which is important for achieving high conversion efficiency. Concentration from 1-sun to 500-sun raises the efficiency of proposed GaSb/GaAs QD IB solar cell from 30% to 50%. We showed that the clue to understanding of performance of GaSb/GaAs QD IB solar cells is the accumulation of charge in QDs and spacers.

An important advantage of proposed GaSb/GaAs QD IB solar cell is that QDs are of type-II and the fact that QDs are located outside of depletion region, at far enough distance so that they cannot cause current leakage through the depletion region. Thermal assistance or solar photons must be involved for confined electrons to escape from such QDs.

Recombination through QDs is a major factor that limits efficiency of solar cells based on QDs buried within the depletion region. Our proposal will help to solve this problem.

## ACKNOWLEDGEMENTS

A. Kechiantz and A. Afanasev acknowledge support from The George Washington University. J.-L. Lazzari acknowledges support from the bilateral Armenian/French project, SCS/CNRS contract #IE-013/ #23545.

## REFERENCES

[1] Luque A., Marty A., and Stanly C., "Understanding intermediate-band solar cells", Nature Photonics 6, 146-152, (2012)
[2] Kechiantz A. M., Afanasev A., and Lazzari J.-L., "Efficiency limit of $Al_xGa_{1-x}As$ solar cell modified by $Al_yGa_{1-y}Sb$ quantum dot intermediate band embedded outside of the depletion region", Photovoltaic Tech. Conf.: Thin Film & Advanced Silicon Solutions, Aix-en-Provence, France (June 6-8, 2012)
[3] Luque A., and Marti A., "Increasing the Efficiency of Ideal Solar Cells by Photon Induced Transitions at Intermediate Levels", Phys. Rev. Lett. 78(26), 5014-5017 (1997)
[4] Luque A., and Marti A., "The Intermediate Band Solar Cell: Progress Toward the Realization of an Attractive Concept", Adv. Matter 22, 160-174 (2010)
[5] Linares P.G., Martı A., Antolın E., Farmer C.D., Ramiro I., Stanley C.R., and Luque A., "Voltage recovery in intermediate band solar cells, Sol. Energy Mater. Sol. Cells 98, 240-244 (2012)
[6] Luque A., Linares P. G., Antolın E., Ramiro I., Farmer C. D., Hernandez E., Tobıas I., Stanley C. R., and Martı A., "Understanding the operation of quantum dot intermediate band solar cells, J. Appl. Phys. 111, 044502 (2012)
[7] Kechiantz A.M., Kocharyan L.M., Kechiyants H.M., "Band alignment and conversion efficiency in Si/Ge type-II quantum dot intermediate band solar cells", Nanotechnology 18, 405401 (2007)
[8] Liang B., Lin A., Pavarelli N., Reyner C., Tatebayashi J., Nunna K., He J., Ochalski T.J., Huyet G., and Huffaker D.L., "GaSb/GaAs type-II quantum dots grown by droplet epitaxy", Nanotechnology 20, 455604 (2009)
[9] Kechiantz A. M., Afanasev A., Lazzari J.-L., A. Bhouri, Y. Cuminal, and P. Christol, "Efficiency limit of $Al_xGa_{1-x}As$ solar cell modified by $Al_yGa_{1-y}Sb$ quantum dot intermediate band embedded outside of the depletion region", Proc. 27$^{th}$ EU PVSEC 16877, 412-417 (2012)
[10] Geller M., Kapteyn C., Müller-Kirsch L., Heitz R., and Bimberg D., 450 meV hole localization in GaSb/GaAs quantum dots, Appl. Phys. Lett. 82, 2706-2708 (2003).